\documentclass[fleqn,usenatbib]{mnras}

\usepackage{newtxtext,newtxmath}

\usepackage[T1]{fontenc}

\DeclareRobustCommand{\VAN}[3]{#2}
\let\VANthebibliography\thebibliography
\def\thebibliography{\DeclareRobustCommand{\VAN}[3]{##3}\VANthebibliography}

\usepackage{graphicx}	
\usepackage{amsmath}	
\usepackage{pdflscape}  

\PassOptionsToPackage{normalem}{ulem}
\usepackage{ulem}

\usepackage{xcolor}
\providecolor{added}{HTML}{0077c0}
\providecolor{deleted}{HTML}{e23e57}

\newcommand{\secref}[1]{\S\ref{#1}}

\title[Dual polarization MWA beam measurements]{Dual Polarization Measurements of MWA Beampatterns at 137 MHz}

\author[A. Chokshi et al.]{A. Chokshi,$^{1,2,3}$\thanks{E-mail:\href{mailto:achokshi@student.unimelb.edu.au}{achokshi@student.unimelb.edu.au}}
J. L. B. Line,$^{2,4}$
N. Barry,$^{1,2}$
D. Ung,$^{4}$
D. Kenney,$^{4}$
A. McPhail,$^{5}$
\newauthor
A. Williams,$^{5}$
R. L. Webster$^{1,2}$
\\
$^{1}$The University of Melbourne, School of Physics, Parkville, VIC 3010, Australia\\
$^{2}$ARC Centre of Excellence for All Sky Astrophysics in 3 Dimensions (ASTRO 3D)\\
$^{3}$CSIRO Astronomy and Space Science (CASS), PO Box 76, Epping, NSW 1710, Australia\\
$^{4}$International Centre for Radio Astronomy Research, Curtin University, Perth, WA 6845, Australia\\
$^{5}$Curtin Institute of Radio Astronomy, GPO Box U1987, Perth, WA 6845, Australia
}

\date{Accepted XXX. Received YYY; in original form ZZZ}

\pubyear{2020}

\begin{document}
\label{firstpage}
\pagerange{\pageref{firstpage}--\pageref{lastpage}}
\maketitle


\begin{abstract}
The wide adoption of low-frequency radio interferometers as a tool for deeper and higher resolution astronomical observations has revolutionized radio astronomy. Despite their construction from static, relatively simple dipoles, the sheer number of distinct elements introduces new, complicated instrumental effects. Their necessary remote locations exacerbate failure rates, while electronic interactions between the many adjacent receiving elements can lead to non-trivial instrumental effects. The Murchison Widefield Array (MWA) employs phased array antenna elements (tiles), which improve collecting area at the expense of complex beam shapes. Advanced electromagnetic simulations have produced the Fully Embedded Element (FEE) simulated beam model which has been highly successful in describing the ideal beam response of MWA antennas. This work focuses on the relatively unexplored aspect of various in-situ, environmental perturbations to beam models and represents the first large-scale, in-situ, all-sky measurement of MWA beam shapes at multiple polarizations and pointings. Our satellite based beam measurement approach enables all-sky beam response measurements with a dynamic range of $\sim$ 50 dB, at 137 MHz.

\end{abstract}




\section{Introduction}

The pursuit for deeper and higher resolution astronomical observations has led to the adoption of low-frequency radio interferometer arrays. Large numbers of relatively simple dipoles, coherently synthesized together, have angular resolutions capable of exceeding the largest traditional dish telescopes. Notably, some of the largest interferometers are now the size of the Earth and beyond. These instruments are ideal for investigations from the local to the early universe. Unfortunately, the spectral windows relevant to such observations are often contaminated by radio frequency interference (RFI) from FM radio, television, and other man-made sources, necessitating that these sensitive instruments be located at some of the most remote and least populous regions of the world.

Electronic interactions between the large number of identical and adjacent elements in an interferometer can lead to complex instrumental responses, exacerbated by disproportionate dipole failure rates due to harsh environmental conditions. This underpins the importance of accurate instrumental beam models which will enable precise calibration of data and increase the sensitivity of various science cases.

The Murchison Widefield Array \citep[MWA\footnote{\url{http://www.mwatelescope.org}};][]{Tingay_MWA_2013, Wayth_MWA_2018} is a precursor to the Square Kilometer Array (SKA\footnote{\url{https://www.skatelescope.org}}), located at the Murchison Radio-astronomy Observatory, in the remote western Australian outback. Designed to observe the low-frequency radio sky between 80 and 300 MHz, one of the MWA's key science goals is detection of redshifted 21 cm emission from the Epoch of Reionization (EoR) \citep{Bowman_MWA_2013, Beardsley_MWA_2019}. In this work we will explore the in-situ measurement of MWA beam shapes, broadly in the context of EoR science.


The high dynamic range of EoR experiments, coupled with the intrinsic chromatic nature of radio interferometers can introduce spectral structure variations, leading to calibration errors which must be constrained to extremely precise levels \citep[e.g.][]{Barry_Calibration_2016, Trott_Wayth_2016, Patil_2017}. The Fully Embedded Element (FEE) beam model \citep{Sutinjo_FEE_2015, Sokolowski_FEE_2017} is a cutting-edge electromagnetic simulation of the tile response using the FEKO\footnote{\url{http://www.feko.info}} simulation package which can be used in EoR pipelines such as the RTS and FHD \citep{Mitchell_RTS_2008, Sullivan_FHD_2012, Barry_FHD_epp_2019}. While accurate simulations of the instrumental beam has been crucial in improving calibration, simulations reflect ideal conditions, which often do not perfectly represent the in-situ reality. This is especially true for the MWA, located in a remote harsh desert, where multiple environmental factors may perturb instrumental beams from their ideal behaviour.

A relatively unexplored aspect of the calibration of radio interferometric data is the instrumental beam. Errors in beam models can introduce flux calibration and polarization errors which may significantly impede the detection of the EoR signal. Simulations by \citet{Joseph_beam_2019} show that beam deformations due to broken dipoles can introduce biases in the 2D power spectrum (PS) up to two orders of magnitude above the expected EoR signal. Laboratory measurements and simulations by \citet{Neben_beamformer_2016} reveal that inter-tile beam variation due to beamformer errors make foreground subtraction infeasible. This is not an insurmountable issue for telescopes which plan on utilizing a foreground avoidance approach, as it is shown that beamformer errors do not contribute significant spectral structure into the theoretically foreground-free regions of the power spectrum.

Spectral features of $\sim10^{-5}$ in the antenna or receiver system can hinder the detection of the EoR signal \citep{Barry_Calibration_2016}. It is possible that spectral structure of this scale could be introduced via errors in beam calibrations. Environmental effects can be large contributors to beam distortions and it is unclear in precisely what ways these distortions contribute spectral structure to the PS, emphasising the requirement for exceptionally well characterised individual beam models for more sophisticated analysis of EoR data.

This paper is presented in the context of EoR science, but has broad implications and the scope to significantly inform a wide variety of science cases which utilise data from wide-field radio interferometers. For example, radio polarimetry studies using the MWA have found significant flux leakage from Stokes I into other Stokes parameters \citep{Bernardi_leakage_2013, Lenc_leakage_2017, Lenc_leakage_2018}. For a Zenith pointings (-27$^\circ$ declination) leakage was $\sim1\%$ and $\sim4\%$ at the edge of the primary beam, increasing to a range of 12-40$\%$ at off zenith pointings. The GaLactic and Extragalaxtic All-sky MWA (GLEAM) survey \citep{Wayth_GLEAM_2015, Hurley-Walker_leakage_2014, Hurley-Walker_leakage_2017} found beam errors to cause frequency and declination dependant errors in Stokes I. Surveys such as GLEAM form the basis for calibration of EoR observations, making a correct flux scale essential. The increasing unreliability of the beam model, away from zenith, causes surveys such as GLEAM to only use the central half-power portion of the primary beam. Accurate beam models would enable the use of a larger portion of the beam with confidence, presenting the opportunity for a significant increase in sensitivity and thus faster experiments and better utilisation of precious telescope time.

A traditional methods of beam measurement, known as radio holography, utilises drift scans of celestial sources of known flux densities to probe cross-sectional slices of the primary beam \citep[e.g.][]{Nunhokee_HERA_beam_2020, Berger_CHIME_2016, Pober_DRIFT_SCAN_2012, Thyagarajan_2011, Bowman_Beam_2007}. Pulsar holography has been proposed to improve polarised beam measurements \citep{Newburgh_CHIME_2014}. A significant impediment to such methods is the faint nature of celestial sources which often have insufficient flux to probe the depths of the beam sidelobes and nulls, especially given that wide-field instruments such as the MWA are sensitive to the whole sky.

An alternate method being explored is the use of radio transmitters mounted on commercially available drones \citep[e.g.][]{ Jacobs_drones_2017, Chang_drone_2015}. This technique has been used as an in-situ validation of two SKA-Low prototype arrays \citep[][]{Paonessa_drones_ska_2020} and LOFAR\footnote{\url{http://www.lofar.org}} antennas \citep[e.g.][]{Ninni_drones_lofar_2020, Bolli_drones_lofar_2018, Virone_drones_lofar_2014}. A distinct advantage of this approach is the control and repeatability of drone flight paths, at multiple frequencies, enabling broadband characterization of beam shapes. While promising, this method comes with a set of drawbacks. Drones have limited altitude ranges and thus operate in the near-field of the instrument as opposed to astronomical observations which occur in the far-field. This is particularly relevant to wide-field instruments, where the projection of drone mounted transmitter beam couples to the Antenna Under Test (AUT) beam, exacerbated as the drone moves further from zenith. Finally, the use of bright radio receivers at radio-quiet zones make such methods challenging for large interferometric arrays such as the MWA, LOFAR, HERA\footnote{\url{https://reionization.org}} and the upcoming SKA-Low.

The final method used to measure beam shapes and the focus of this paper, utilises satellites as bright radio sources with known trajectories, to probe cross-sectional slices of the AUT. Advantages of this method include: bright satellites enabling high dynamic range observations of the beam and sidelobes; sources emitting in the far field; the precision of orbital tracks creating new slices of the AUT beam with each orbit. This method was neatly demonstrated by \citet{Neben_ORBCOMM_2015} using a test MWA tile and by \citet{Neben_HERA_2016} using a prototype HERA dish at the NRAO Green Bank Observatory\footnote{\url{https://greenbankobservatory.org}}. The work of \citet{Line_ORBCOMM_2018} represents the first in-situ demonstration of this method at the MWA site.

This paper represents the first large-scale, in-situ, all-sky measurement of MWA beam shapes at multiple polarizations and pointings, with the aim to quantify inter-tile variations and measure environmental beam distortions at 137 MHz using communication satellites. Our methodology demonstrates a passive parallel monitoring system, which measures the beam shapes of MWA tiles in parallel to regular observations with no disruption to the operation of the telescope. As this setup was built using cheap off-the-shelf components, and the analysis is carried out using our open-source \texttt{python}\footnote{\url{https://www.python.org}} package called \texttt{EMBERS}\footnote{\url{https://embers.readthedocs.io}} (Experimental Measurement of BEam Responses with Satellites) \citep{Chokshi_EMBERS_2020}, we present it as a prime candidate for a passive beam monitoring system for large interferometric arrays such as the MWA, HERA, LOFAR and SKA-Low. As Radio Frequency Interference (RFI) encroaches on the last remaining radio-quiet observatories, archival data becomes ever more valuable. The addition of measured beam shapes could be critical to the analysis of this data in the future, when more sophisticated analysis techniques are developed.

A description of the MWA Telescope, experimental setup and data acquisition system are explained in \secref{sec:expt}, following which \secref{sec:data} outlines our data analysis method. In \secref{sec:results} we present challenges encountered in our analysis, experimental biases and the results. Finally, in \secref{sec:conclusions}, we discuss implications of this work and possible future directions.

\section{Experimental Method}
\label{sec:expt}

The approach taken in this experiment is an extension of investigations presented in \citet{Line_ORBCOMM_2018} and \citet{Neben_ORBCOMM_2015}. The premise of this work is based around using radio satellites, with well known orbital trajectories, to probe the beam response of MWA tiles. The power received by the Antenna Under Test (AUT) is the product of the beam response $B_\mathrm{AUT}$ and the flux transmitted by the satellite $F$. A reference antenna with a simple, well known beam response $B_\mathrm{ref}$ is used to record the modulation of the transmitted flux, and can subsequently be used to compute the beam shape of the AUT. The power received by the AUT and reference antenna are $P_\mathrm{AUT} = B_\mathrm{AUT}F$ and $P_\mathrm{ref} = B_\mathrm{ref}F$ respectively. These expression can be reduced to give us the response of the AUT, described by:
\begin{equation}
    B_\mathrm{AUT} = \frac{P_\mathrm{AUT}}{P_\mathrm{ref}}B_\mathrm{ref}.
	\label{eq:beam_eq}
\end{equation}
With each satellite pass, we measure a cross sectional slice of the AUT beam response. With sufficient observation time, an all-sky beam response is built up.

\subsection{The Murchison Widefield Array}
\label{ssec:mwa}

The MWA is an aperture array telescope, with 128 receiving elements or tiles, each constructed from a grid of $4\times4$ dual polarization bow-tie dipoles, mounted on a $5\times5$ m reflective metal mesh \citep{Tingay_MWA_2013}. The two orthogonal linear polarizations of the MWA tiles are labled XX and YY, with dipoles aligned along the East-West and North-South directions respectively. MWA tiles have a wide field of view, with a full-width half-maximum $\sim 25^\circ$ at 150 MHz, which can be steered using an analogue delay-line beamformer. The beamformers have a set of quantised delays available, which results in a set of 197 discrete pointings to which the beamformer can point the phase-center of the MWA beam. The phased array design of MWA tiles improves the collecting area of tiles, at the expense of additional complexity introduced to the beam shapes.

\subsection{Data acquisition}
The experimental setup used in this work is based on \citet{Line_ORBCOMM_2018} and expanded to accommodate our new science goals which differ from previous methods in a few key ways. 

\begin{figure}
	\includegraphics[width=\columnwidth]{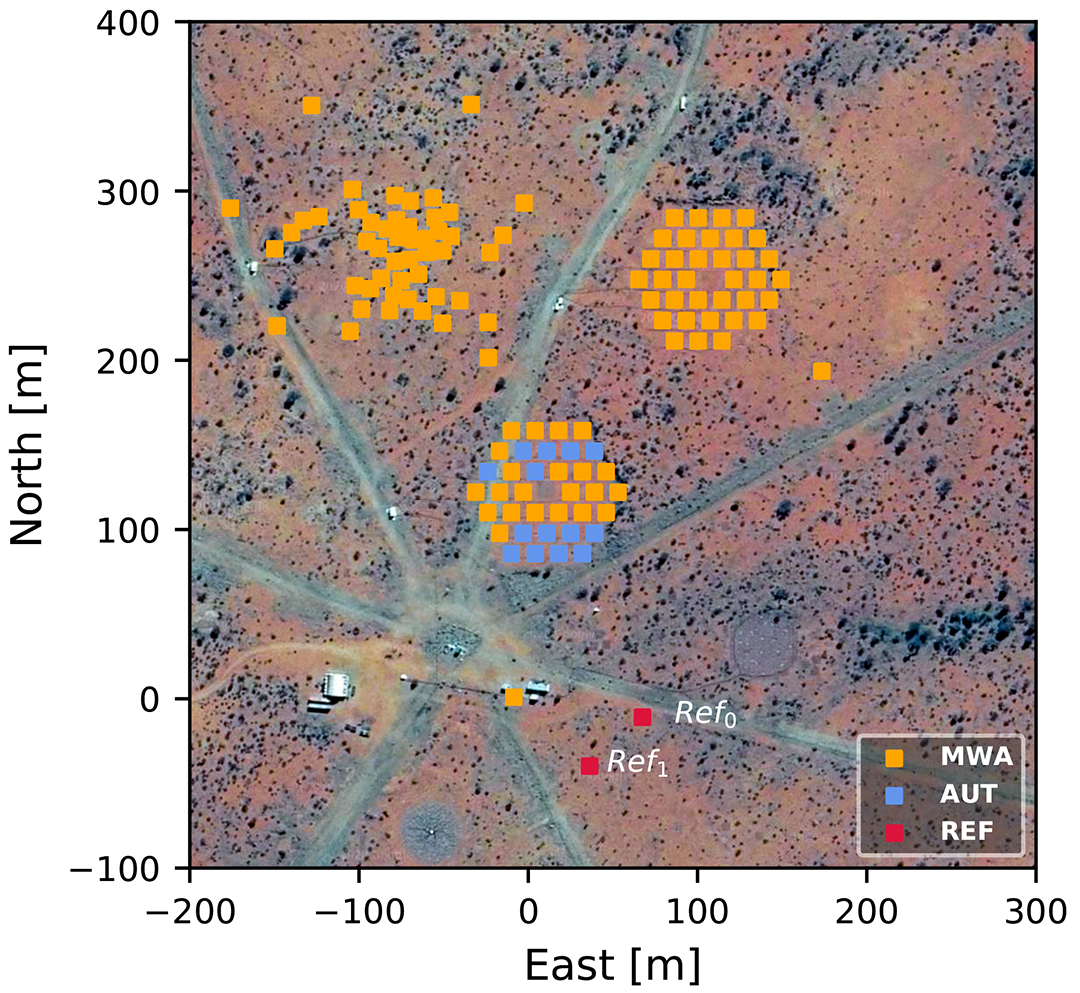}
    \caption{The positions of the AUTs (blue) and the Reference antennas (red). The ochre points represent the rest of the compact core of the MWA.}
    \label{fig:mwa_map}
\end{figure}

We measure the all-sky beam response of 14 MWA tiles, over a 6 month period, at both instrumental polarizations (XX, YY) and at multiple pointings using ORBCOMM\footnote{\url{https://www.orbcomm.com/en/networks/satellite}} communication, METEOR\footnote{\url{http://www.russianspaceweb.com/meteor-m.html}} and NOAA\footnote{\url{https://www.noaa.gov/satellites}} weather satellites. This work is the first demonstration of parallel, in-situ beam measurements without disruption to the telescope's observational schedule. The 14 MWA tiles are a part of the inner core of the compact configuration of the MWA array, located within the ``Southern Hex'' as shown in Figure \ref{fig:mwa_map}. The names of the tiles can be found in Table \ref{tab:tiles}. In addition to the Zenith pointing of the telescope, measurements of the beam response are carried out at two off-zenith pointings (see Table \ref{tab:pointings}).

\begin{table}
	\caption{Reference and MWA tiles used in the Experiment. Each tile is dual-polarised with both XX and YY dipoles. For example, the rf0 tile has rf0XX and rf0YY arrays.}
	\label{tab:tiles}
	\centering
	\begin{tabular}{llllllll}
		\hline
		rf0 & rf1 & S06 & S07 & S08 & S09 & S10 & S12 \\
		S29 & S30 & S31 & S32 & S33 & S34 & S35 & S36 \\
		\hline
	\end{tabular}
\end{table}

Radio frequency (RF) signals are simultaneously recorded from 14 MWA tiles and two reference antennas, in both XX and YY polarizations. The reference antennas are constructed using a single dual polarization MWA dipole, centered on a $5\times5\,$m conductive ground mesh. Custom-built RFI shielded circuits are used to power the Low Noise Amplifiers (LNAs) within the dipole, and retrieve data via coaxial cables. These field boxes contain secondary LNAs, to further amplify RF signals, and Bias-Ts which facilitate data and power transfer through coaxial cables. These field boxes are placed near the reference antennas and are connected with long coaxial cables, to RF Explorers\footnote{\url{http://rfexplorer.com}} located within a RFI shielded hut approximately 50 m away. RF signals from the tiles are acquired inside the MWA receivers \citep*[see][]{Tingay_MWA_2013}, using beam splitters, after amplification and filtering by the Analogue Signal Conditioning unit. These are passed to RF Explorers installed within the receivers. 

\begin{figure*}
	\includegraphics[width=\textwidth]{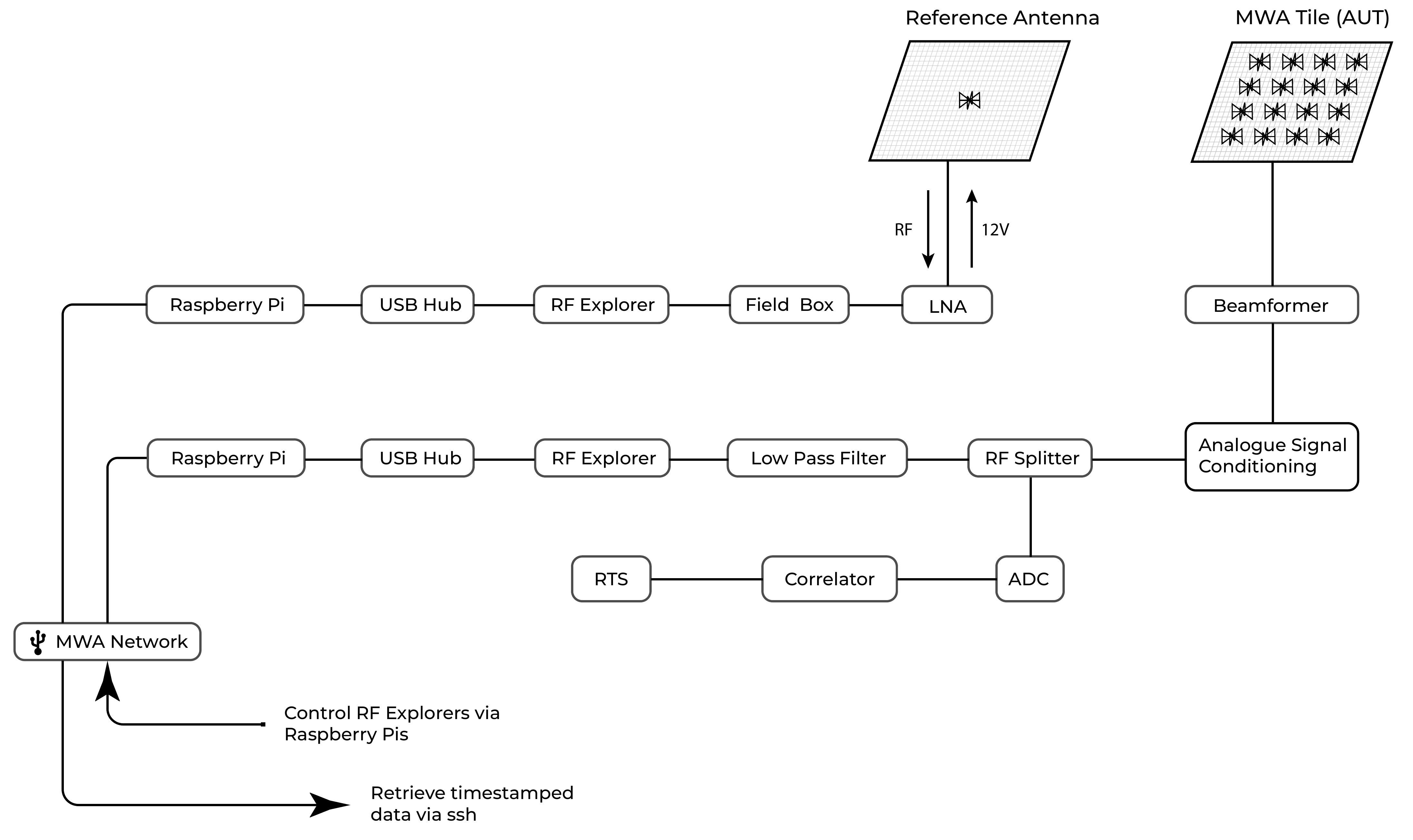}
    \caption{Flow chart of our experimental setup to measure MWA beam shapes. \textbf{Top:} The reference dipole receive satellite signals which is amplified by a Low Noise Amplifier (LNA). A Bias-T in the field box supplies the LNA with a 12V power supply and transmits the satellite signal from the dipole to the field box. Long coaxial cables carry the amplified signal to a RFI-shielded hut for analysis by a RF Explorer, the results of which are saved by a Raspberry Pi computer. \textbf{Bottom:} RF signals received by the Antenna Under Test (MWA Tile) are fed to an analogue beamformer, which introduces time delays to the signals from the 16 dipoles corresponding to the pointing of the telescope. The signals are combined and transmitted via long coaxial cables to an MWA receiver. Within the receiver, the Analogue Signal Conditioning unit performs amplification and filtering before passing it to a signal splitter. The splitter sends half the signal on its usual path to the correlator, while the other half passes through a low-pass filter before being analysed by a RF Explorer and saved by a Raspberry Pi. The USB hubs supply power to the RF Explorers and facilitates the transfer of data from multiple RF Explorers to the Raspberry Pi. The Raspberry Pis are connected to the MWA network, from which they can be remotely controlled and transfer data.}
    \label{fig:setup}
\end{figure*}

The RF Explorers are set to have a spectral resolution of 12.5 kHz, sampling 112 frequency channels between 137.150 MHz and 138.550 MHz. This frequency window was chosen to observe Meteor and NOAA weather satellites and the ORBCOMM constellation of communication satellites, which provide excellent sky coverage. The signal is acquired at a rate between 6 - 9 samples per second, limited by the hardware in the RF Explorers. A set of five Raspberry Pi\footnote{\url{https://www.raspberrypi.org}} single-board computers are used to control and retrieve data from the 32 RF explorers connected to 14 MWA tiles and 2 reference antennas. The positions of the antennas can be seen in Figure \ref{fig:mwa_map}. USB hubs are used to power and facilitate the control of multiple RF Explorers by a single Raspberri Pi. An outline of our experimental setup can be found in Figure \ref{fig:setup}.

The Raspberri Pi's are connected to the MWA network via ethernet cables, enabling remote control over the experiment. Network access allows the synchronisation of the Raspberri Pi's by syncing them to the same NTP server. The Raspberri Pi's control the RF Explorers using a custom \texttt{python} script and the \texttt{pySerial}\footnote{\url{https://pythonhosted.org/pyserial}} module. Every 24 hours, a scheduled \texttt{cron job}\footnote{\url{http://man7.org/linux/man-pages/man8/cron.8.html}} transfers the recorded RF data to an external server and, using the \texttt{at}\footnote{\url{http://man7.org/linux/man-pages/man1/at.1p.html}} command, schedules a day of 30 minute observations across all the RF Explorers.

The beam splitters allowed the experiment to run concurrently with normal MWA operations, meaning the pointing of the telescope was dictated by the regular observational schedule. A large amount of data was recorded using this setup, invariably including a significant portion irrelevant to this project. Though the experiment was plagued by technical failures of the RF Explorers, USB hubs and a rare lightning strike, between 12$^{\mathrm{th}}$ September 2019 and 16$^{\mathrm{th}}$ March 2020, over 4000 hours of raw data were collected.

\begin{table}
	\centering
	\caption{MWA beamformer pointings used in this work}
	\label{tab:pointings}
	\begin{tabular}{lllr}
		\hline
		MWA Pointing & Altitude & Azimuth & Integration [h]\\
		\hline
		0 & 90$^\circ$ & 0$^\circ$ & $\sim900$\\
		2 & 83$^\circ$11$'$28.32$''$ & 90$^\circ$ & $\sim350$\\
		4 & 83$^\circ$11$'$28.32$''$ & 270$^\circ$ & $\sim350$\\
		\hline
	\end{tabular}
\end{table}

\section{Data Analysis}
\label{sec:data}

\begin{figure*}
	\includegraphics[width=\textwidth]{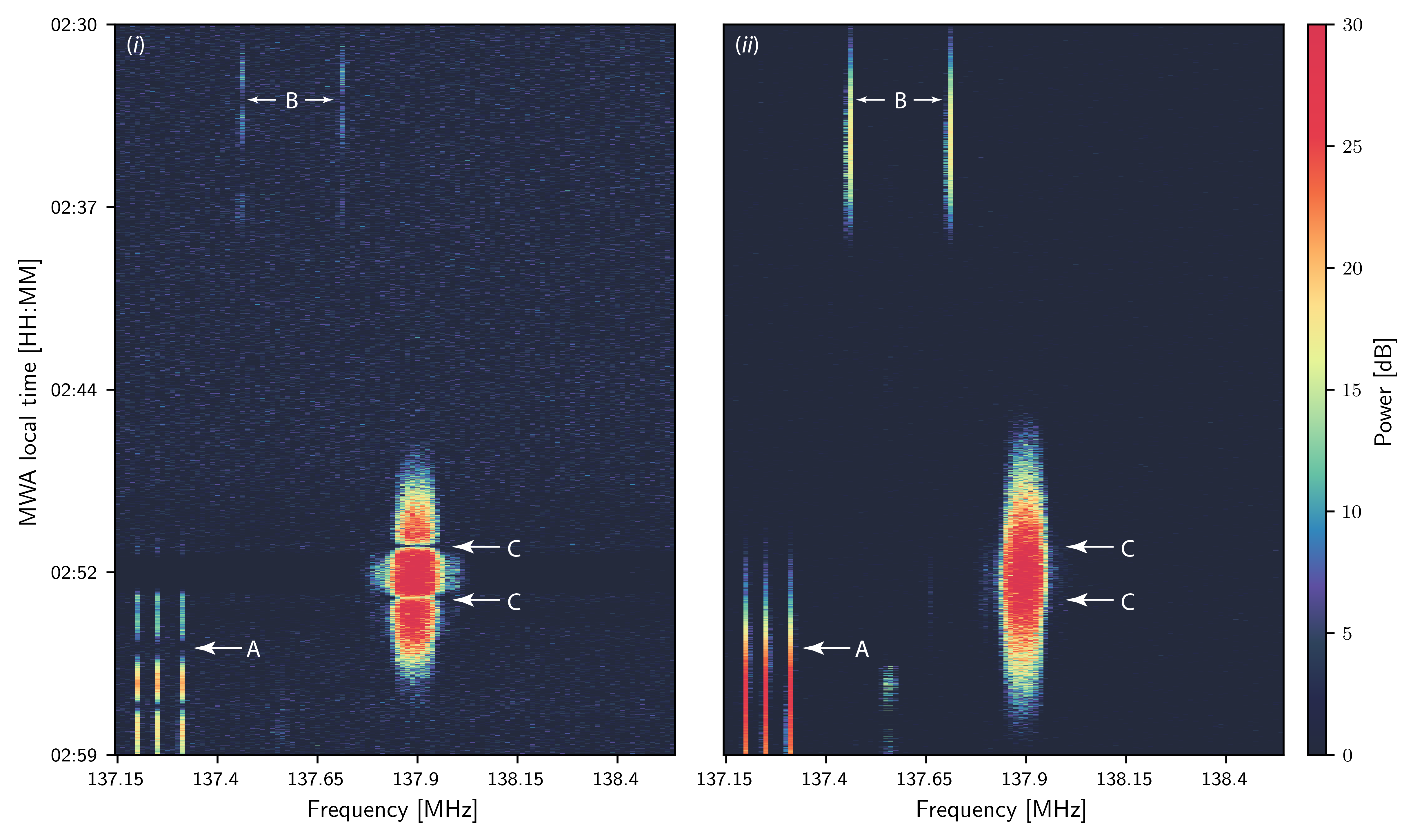}
    \caption{A sample set of raw data observed between 2:30AM and 3:00AM on 10/10/2019. The image on the left (i) is tile S10XX while the image on the right is data from reference ref0XX. Both sets of data have been scaled to have a median power of 0 with a dynamic range of 30dB. Interesting features have been annotated at the same positions in each plot with arrows indicating points of stark differences between the plots. The flux received by the MWA tile (i) drops to zero at the positions of the nulls at the edge of the MWA primary beam, which are absent in the reference antenna (ii). We find that ORBCOMM satellites generally have narrow band transmissions, occupying no more than 2 channels, as seen in A and B. Meteor weather satellites have significantly broader spectral footprints, occupying up to 10 channels, as seen in C.}
    \label{fig:waterfall}
\end{figure*}

A sample of the raw data can be seen in Figure \ref{fig:waterfall}, in the form of a waterfall\footnote{time vs. frequency} plot. ORBCOMM satellites were found to transmit in narrow frequency bands, occupying up to two 12.5 kHz channels. In contrast, weather satellites exhibit a broader spectral signature, occupying up to 10 consecutive channels. In \citet{Neben_ORBCOMM_2015}, an `ORBCOMM user interface box' was used to match satellite ephemerides to transmission frequencies of satellites above the horizon. As this technology is not commercially available, \citet{Line_ORBCOMM_2018} used satellite ephemerides, published by Space-Track.org\footnote{\url{https://www.space-track.org}}, to match satellites above the horizon to observed RF signals seen in waterfall plots similar to Figure \ref{fig:waterfall} and manually create a map of ORBCOMM transmission frequencies.

Multiple ORBCOMM satellites are often above the horizon simultaneously, and are observed to periodically shift transmission frequencies to avoid inter-satellite interference. With observations spanning more than 6 months, and the resulting large volume of data it became infeasible to manually determine the transmission frequencies of every satellite pass. This necessitated the development of an automated system of matching satellite ephemerides and RF data, described in detail in \secref{ssec:ephem}. 

\subsection{Data conditioning}

 Before the analysis of our data can proceed, it must be pre-processed to ensure that sensible comparisons can be drawn between the tiles and references. A complication we encountered was that different RF Explorers recorded the data at different temporal rates, ranging between 6 and 9 Hz.  We attribute this issue to two distinct batches of RF Explorers used. The first batch of 8 were purchased in 2017, and recorded data at a rate between 6-7 samples per second, while the remaining 24 RF Explorers were purchased in 2019 and recorded data at a rate between 8-9 samples per second. Though the model numbers of the RF Explorers and their configuration settings were identical, we infer that there must have been hardware improvements in the more recently manufactured modules.
 
An optimal balance between the RF Explorers sampling rate, Signal-to-Noise and the sky coverage of our selected satellites, determines the N-side of our HEALPix \citep{Gorski_Healpix_2005} maps. We use a N-side of 32, corresponding to an angular resolution of 110 arcmins. An important consideration at this stage was that Satellites in Low Earth Orbit typically transit the visible sky in 5-15 minutes, depending on their orbital altitudes \citep{Cakaj_LEO_2009}. Typical transit periods of satellites used in this experiment were observed to be in the 15 minute range, at which satellites took $\sim$9 seconds to transit across one 110 arcmin HEALPix pixel.
 
\begin{figure}
	\includegraphics[width=\columnwidth]{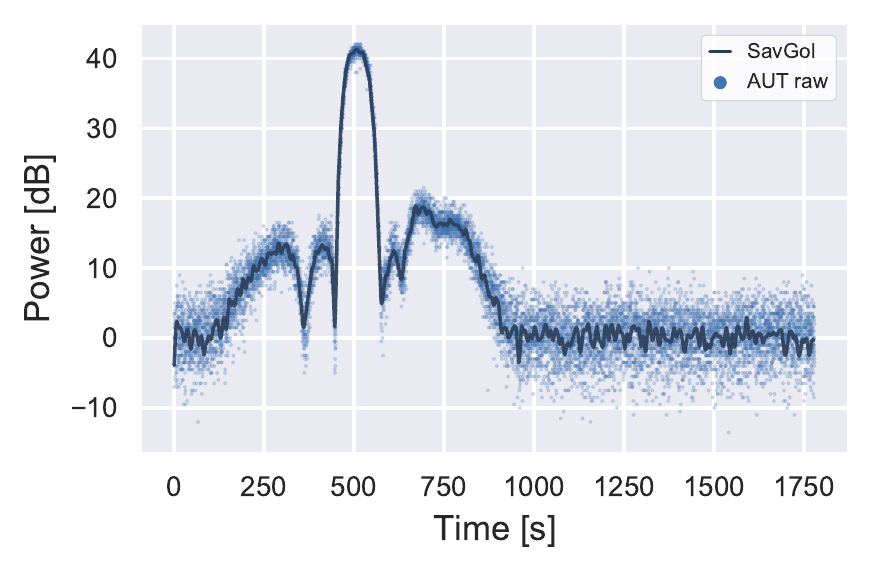}
    \caption{A single channel of raw data with a bright satellite pass. The solid curve is the SavGol smoothed data, interpolated down to 1 Hz.}
    \label{fig:savgol}
\end{figure}

 The calculation above indicates that the raw data is highly oversampled, providing a certain leeway to get around the issue of varied temporal sampling. An iterative Savitzky–Golay (SavGol) filter is selected to smooth the raw noisy data, while preserving it's high dynamic range. Initially, a SavGol filter with a small window is used to preserve the depth of the null in the beam response, followed by a second SavGol filter with a larger window to smooth short time-scale noise present in the data. We then interpolate our data down to a 1 Hz frequency, while retaining multiple data points per HEALPix pixel. This enables us to compare our tile and reference data accurately. Figure \ref{fig:savgol} shows the concurrence between the raw data and the SavGol smoothed, interpolated data. 

 A noise threshold is defined at this stage, allowing further analysis to be limited to RF satellite signals above the noise floor. In a 30 minute observation, typically 3-7 of the 112 frequency channels contain satellite signals. These channels are identified to first order by having peak signals above a single standard deviation ($\sigma_\mathrm{raw}$) of the data. Excluding these occupied channels, the Median Absolute Deviation (MAD\footnote{MAD - robust statistic more resilient to outliers than standard deviation})  $\sigma_\mathrm{noise}$ and the median $\mu_\mathrm{noise}$ of the remaining noisy channels are used to define a noise threshold shown in Equation \ref{eq:noise}. The noise threshold for both $P_\mathrm{AUT}$ and $P_\mathrm{ref}$ is computed for every time-step with 
\begin{equation}
    P_\mathrm{noise}=\mu_\mathrm{noise} + \sigma_\mathrm{noise}.
	\label{eq:noise}
\end{equation}

\noindent If $P<P_\mathrm{noise}$ for either the AUT or reference data, the time-step data is flagged as ``noisy''.

 As observations were carried out in parallel to the regular observational schedule of the MWA, satellite RF data is recorded at all pointings the telescope visited over the course of the experiment, accumulating over 4000 hours of raw data. The total integrated data at most pointings fell far short of the $\sim 400$ hours required to make maps with a N-side of 32. The Zenith and two EoR 2, 4 pointings met this criteria, resulting in $\sim 1600$ hours of usable data. The data is sorted based on pointing, and separate maps are created for each. See Table \ref{tab:pointings} for details about the pointings and the amount of usable data collected at each.

\subsection{Satellite ephemerides}
\label{ssec:ephem}

Satellites transmit data to Earth on their allocated ``downlink'' frequency, the exact location of which are often proprietary. Reliable sources of data regarding spectrum allocations in the 137-138 MHz band are scarce and often outdated. Our initial estimate of $\sim70$ active satellites within our window was optimistic, with a total of 18 satellites being regularly observed in our data. Some of our initial satellite candidates were no longer actively transmitting, while others presumably transmitting marginally outside our frequency window. Table \ref{tab:sats} contains information on the satellites we used. The orbital parameters (ephemeris) of most satellites are recorded multiple times a day by USSPACECOM and are published by Space-Track.org. The ephemerides of our satellites are downloaded in the form of Two Line Elements\footnote{\url{http://www.satobs.org/element.html}} (TLEs). A custom \texttt{python} script reads these TLEs and accurately (within $\sim$10 arcsec at epoch) computes when the satellites are above the horizon and their trajectories in the sky, at the MWA telescope . The  \texttt{Skyfield}\footnote{\url{https://rhodesmill.org/skyfield}}\citep{skyfield} software package was instrumental to these calculations.

\begin{table}
	\caption{Satellite constellations and frequency bands. }
	\label{tab:sats}
	\begin{tabular}{lll}
		\hline
		Constellation & Spectral band [MHz] &  Satellites observed\\
		\hline
		ORBCOMM & 137.2 -137.800    & 15\\
		NOAA    & 137.1 -137.975    &  2\\
		METEOR  & 137.1 -137.975    &  1\\
		\hline
	\end{tabular}
\end{table}

\subsection{Frequency mapping}
\label{ssec:freq_map}

The sheer quantity of data made it infeasible to manually determine transmission frequencies of our satellites. Instead, we developed a method to automatically cross match satellite ephemerides and raw RF data, identifying the transmission frequency of every satellite in each 30 minute observation. Using the ephemerides of each satellite, a temporal window within the RF data is identified, within which transmissions are expected to be found. We define a set of criteria to identify the correct frequency channel. 

\begin{description}
  \item[\textbf{Window Occupancy:}] $W_\mathrm{RF}$ is the percentage of RF signal above the noise threshold $P_\mathrm{noise}$ (Eq. \ref{eq:noise}), within the temporal window. Identified satellites were required to have an occupancy in the range $80\% \leqslant W_\mathrm{RF} < 100\%$. The lower limit accounts for satellite passes close to the horizon, where long noise-like tails are observed on either end of the satellite data.
  \item[\textbf{Power Threshold:}] $P_\mathrm{peak}$ is introduced to set a minimum peak satellite power. It was observed that channels adjacent to a bright satellite pass were often observed to be contaminated with lower power, noise-like, RF signals. This probably occur due to transmission bandwidth marginally exceeding the 12.5 kHz channel width of the RF Explorers, leading to spectral leakage. Such channels typically have peak power in the range of 10-15 dB, compared to the 20-40 dB peak powers. To eliminate these contaminated channels, we require identified channels to have $P_\mathrm{peak} \geqslant 15 \mathrm{dB}$.
  \item[\textbf{Triplets:}] It is common to observe pairs or triplets of almost identical signal, as seen in labels A, B of Figure \ref{fig:waterfall}, which often pass both filters described above. In such cases, the channel with the higher window occupancy is selected, indicative of a superior match between RF data and satellite ephemerides.
\end{description}

While this method has been highly effective, it is not foolproof. At later stages in the analysis, described in Section \ref{ssec:tile-map}, obvious errors in this method are eliminated by implementing a goodness of fit test between measured beam profile and the FEE simulated model.

\subsection{Map making}
\label{ssec:feko}

FEKO simulations were run to create simulated beam models of the reference antennas $B_\mathrm{ref}$, using on-site measurements of the ground screen and dipole positions, identical to those used to generate the FEE models of the MWA beam. These models are used to make maps of the tile responses by computing $P_\mathrm{AUT}/P_\mathrm{ref} \times B_\mathrm{ref}$ (Eq. \ref{eq:beam_eq}) for each satellite pass, for every pair of AUT and Reference. The different amplifications that the RF signals undergo along the two distinct signal paths -- through the field box for the reference data, and via the beamformer and the Analogue Signal Conditioning unit for the AUTs -- have not been considered (see Figure \ref{fig:setup}). Since we are interested in the profile of the signal, rather than the absolute power, a least-squares method is used to assign each satellite pass, a single multiplicative gain factor $G_\mathrm{FEE}$, effectively fitting it to power level of a corresponding slice of the FEE model.

This slice of the beam response is now projected onto a HEALPix map with a N-side of 32, using the satellite ephemerides, resulting in a map with an angular resolution of 110 arcmins, a good balance between integration per pixel and resolution. Each pixel of the map now contains a distribution of values from multiple satellite passes, the median of which gives us a good estimate of the beam response, without being influenced by outliers.

\section{Results}
\label{sec:results}

\subsection{Null tests}
\label{ssec:null-test}

\begin{figure}
	\includegraphics[width=\columnwidth]{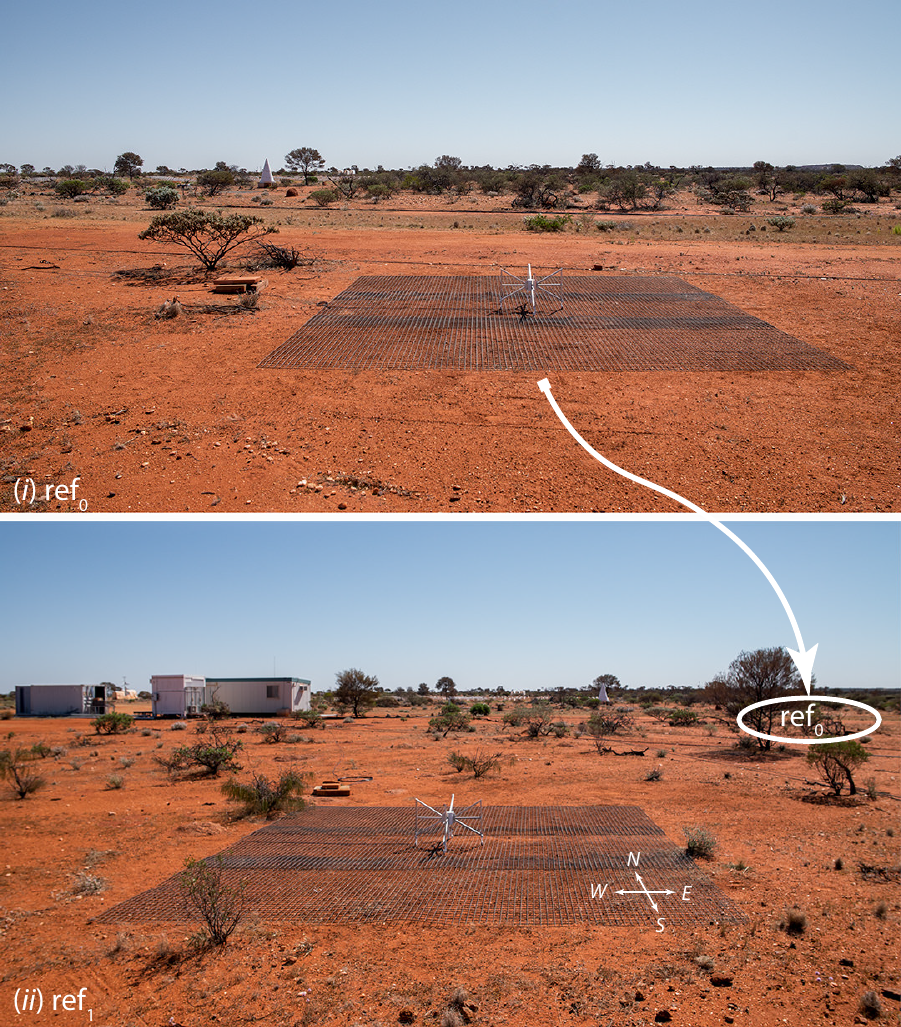}
    \caption{The reference antennas (i): ref$_0$ and (ii): ref$_1$ on site. In the bottom panel, the RFI shielded huts can be seen, as well as the position of ref$_1$ in the distance behind a bush.}
    \label{fig:refs}
\end{figure}

Two reference antennas ref$_0$ and ref$_1$, seen in Figure \ref{fig:refs}, were used in this experiment. This provides the ability to perform a null test to characterize the differences between the beam patterns of the references, and their FEKO simulated models. The ratio of the beam powers, for a set of perfect reference antennas, should ideally be unity and $P_\mathrm{ref0}/P_\mathrm{ref1} = 1$ should hold true for all satellite passes. Deviations from this expression are indicative of systematics such as alignment errors and imperfections in the ground screen, soil, dipole or the surrounding environment.

\begin{landscape}
    \begin{figure}
	    \includegraphics[width=\columnwidth]{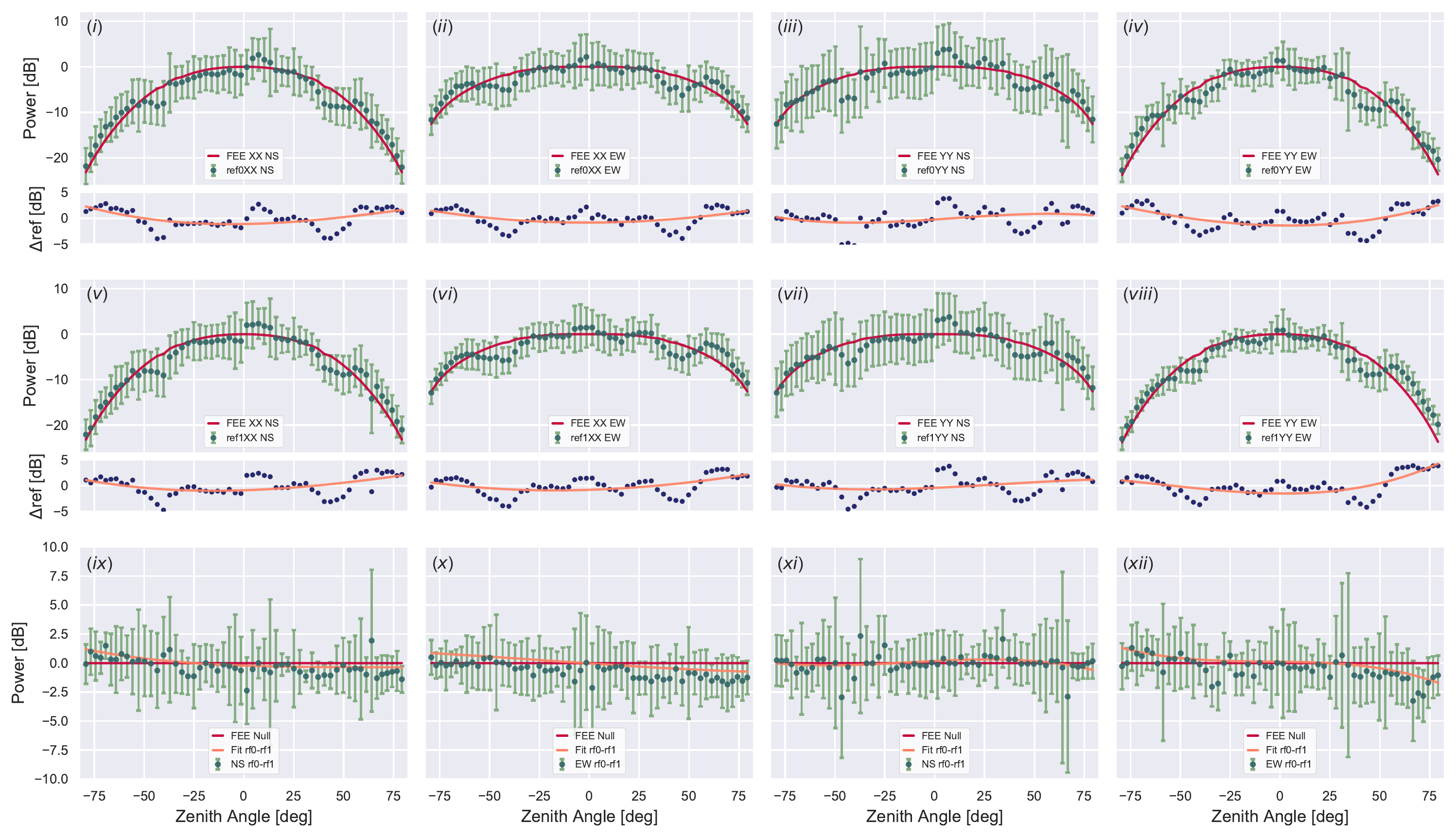}
        \caption{\textbf{Null Test Results:} The first row (i)-(iv) represent slices of HEALPix maps generated from RF data of ref$_0$. (i) and (ii) are North-South (NS) and East-West (EW) slices of the XX polarization of ref$_0$ while (iii) and (iv) are matching NS, EW slices of the YY polarization of ref$_0$. The green data-points indicate the median value of each HEALPix pixel, with the median absolute deviation as the error bars. The crimson curves represent corresponding slices of the FEKO reference model (Section \ref{ssec:feko}). The difference between the data and model $\Delta$ref are plotted in the lower panel as blue points. The orange curve is a third order polynomial fit to the residuals. The second row (v)-(viii) show an identical analysis performed on ref$_1$. The bottom row (ix)-(xii) are the null tests, each computed from the two preceding plots. In (ix), the crimson line represent the the ideal null test while the green data represents the difference between ref$_0$ from (i) and ref$_1$ from (v). The error bars are computed by propagating errors from (i) and (v) while the orange curve shows a third order fit to the null test. (x)-(xii) are similar to (ix), each being calculated from the two plots above it.}
        \label{fig:null-test}
    \end{figure}
\end{landscape}

The results of the null test are shown in Figure \ref{fig:null-test}. The first row (subplots (i)-(iv)) shows slices of the ref$_0$ HEALPix map along both East-West (EW) and North-South (NS) directions for XX and YY polarizations, respectively. The median of the distribution of values in each pixel is power P$_\mathrm{ref0}$, while an estimate of the errors is determined from the Median Absolute Deviation  $\sigma_\mathrm{MAD}$ of the distribution. These are compared to corresponding slices of the reference FEKO model B$_\mathrm{ref}$, and the residuals $\Delta $ref$_0 = $P$_{\mathrm{ref_0}} - $B$_\mathrm{ref}$ are fit with a third order polynomial. The second row (subplots (v)-(viii)) is an identical analysis carried out for the ref$_1$ HEALPix map. The null test is performed in the third row (subplots (ix)-(xii)), where corresponding slices of ref$_0$ and ref$_1$ HEALPix maps are compared. The green data represents a pixel to pixel comparison between of ref$_0$ and ref$_1$, with error bars propagated in quadrature from the $\sigma_\mathrm{MAD}$ of each references. We also compare the fits to the residual power $\Delta $ref (orange curve), seen in the lower panels of the first two rows of Figure. \ref{fig:null-test}. 

\begin{figure}
	\includegraphics[width=\columnwidth]{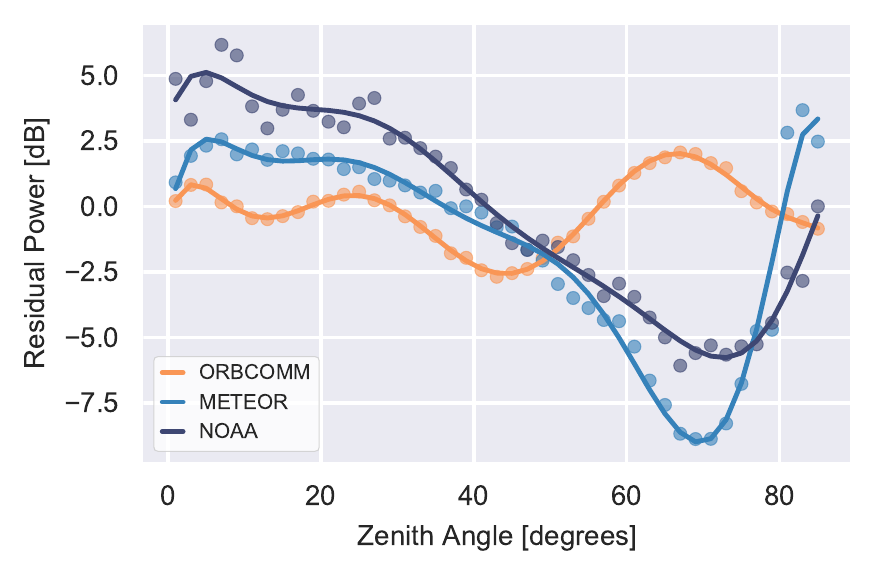}
    \caption{Radially averaged reference residual power, displaying unique beam profiles for each of the three types of satellites used in the analysis, validating our methodology and null tests.}
    \label{fig:residuals}
\end{figure}

An interesting pattern emerges in the residuals between the map slices and FEKO model $\Delta$ref (Figure \ref{fig:null-test} (i)-(viii) lower panels). For zenith angles between $30^\circ - 60^\circ$, a systematic deficit of power is observed with residual power structure observed with deviations up to $\pm2$ dB from the FEKO reference models. This feature is investigated by summing the residuals of all four reference HEALPix maps and averaging the results in $2^\circ$ radial bins. By classifying the data according to the progenitor satellite type, an illuminating pattern emerges, shown in Figure \ref{fig:residuals}. We achieve a good fit of the radial residuals using a 8$^\textrm{th}$ order polynomial. Each satellite has a distinct and well defined residual structure. These residuals represent a profile measurement of the beam shapes of satellite transmitting antennas. This can be understood by considering that satellites primarily focus on transmitting data downwards, normal to the surface of Earth. As satellites rise, the reference antennas observe RF transmitted power convolved with the sidelobes of the satellite beam shapes, which is attenuated away from its primary beam (pointed to the surface).

The amplitude and structure of these residuals may appear significant to our analysis, but are in-fact accounted for by our primary Equation \ref{eq:beam_eq}. This can be illustrated by considering the concurrent measurement of satellite data by both MWA and reference tiles. Any modulation encoded in the data transmitted by the satellites will be identically recorded by all antennas, convolved with individual tile beam shapes. The ratio of observed powers in Eq. \ref{eq:beam_eq} ($P_\mathrm{AUT}/P_\mathrm{ref}$) will neatly divide out any satellite beam structure or modulation encoded within the incoming RF data.  

We note the slightly exaggerated slope in the null test of the EW slice of the YY reference maps, as seen in the last column of Figure. \ref{fig:null-test} (subplots (iv), (viii), (xii)). On further inspection of subplot (viii), we note that the East edge of the ref$_1$ receives $\sim2$ dB less, and the West edge receives $\sim2$ dB more power that the corresponding slice of ref$_0$. We suggest that this discrepancy probably results from a slight EW gradient in ground screen or the dipoles of the tile, which points the bore-sight of the dipole marginally off-zenith.

 The agreement between the fits to the residuals (orange curve in subplots (ix)-(xii) of Figure. \ref{fig:null-test}) and the expected null (red lines) represent a good validation of our experimental procedure described in section \ref{sec:expt}. We observe less than a $\sim$0.5 dB error in the central 25$^\circ$ of the reference model, corresponding to the primary lobe of the MWA beam at 137 MHz. These errors do increase as we move towards the horizon, reaching a maximum of $\sim$2 dB, in our most inaccurate reference. This validates the efficacy of our null test, in characterising systematic effects from the references, which propagate into the beam maps created in the following sections (see grey errorbars in Figure \ref{fig:tile-slices}).  

The null tests display a marginally better performance of reference tile 0 (ref$_0$). Despite this, we have chosen to use reference tile 1 (ref$_1$) in proceeding sections as hardware failures on ref$_0$ resulted in more data and better sky coverage for ref$_1$.

\subsection{RF explorer gain calibration}
\label{ssec:rfe-gain}

\begin{figure}
	\includegraphics[width=\columnwidth]{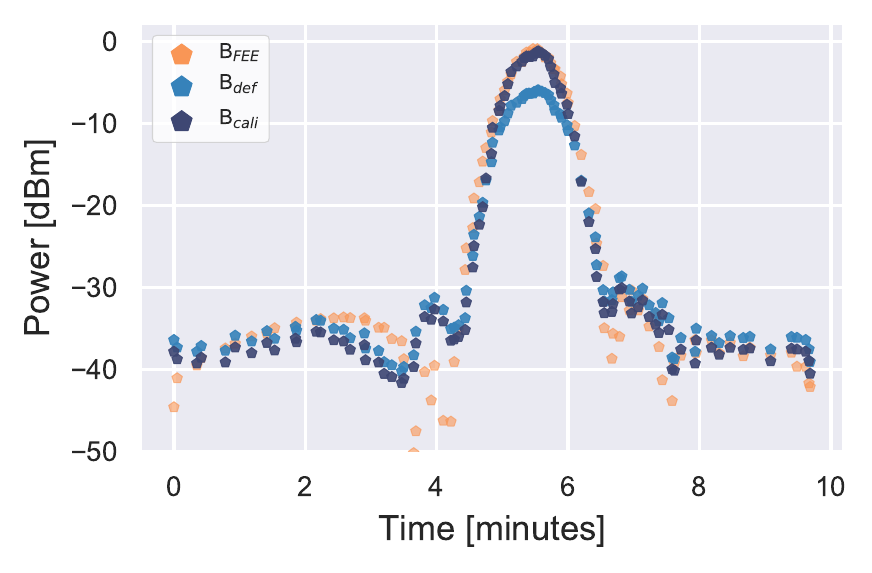}
    \caption{A bright satellite pass  recorded by the non-linear gain of the AUT RF explorer, which results in a deformed beam model B$_\mathrm{def}$ compared to a corresponding slice of the FEE model B$_\mathrm{FEE}$. The power of B$_\mathrm{def}$ is significantly lower than B$_\mathrm{FEE}$, in the primary beam. The efficacy of the RF explorer gain calibration method is demonstrated by the dark blue data points B$_\mathrm{cali}$ which result from applying the gain calibration solution to the distorted beam model (light blue). The nulls of the FEE model extend beyond the depth of the recorded data due to the -50dB sensitivity of the experiment. A significant mismatch between B$_\mathrm{FEE}$ and B$_\mathrm{cali}$ is observed around the 4 minute timestamp. This error can probably be attributed to a combination of a gradient in the ground screen and a slight rotation of the tile, which lead to significant deviations around the edges of the steep nulls as explored in \secref{sec:conclusions}.}
    \label{fig:rfe-gain}
\end{figure}

During the last stages of the experiment, we noticed that the very brightest satellite signals exceeded the maximum recommended power of the RF explorers, resulting in the internal amplifiers entering a non-linear regime. Unfortunately, the limited dynamic range of the RF Explorers coupled with the high dynamic range of satellite observations resulted in almost no leeway for errors in this regard. This effect was only present in RF Explorers recording data from MWA tiles, via the MWA receivers and is apparent in Figure \ref{fig:rfe-gain}, where the light blue raw tile data is $\sim6$ dB lower than a corresponding slice of the FEE model (yellow curve). This deficit of measured power in the primary lobe was unexpected as the primary lobe has been well characterized \citep{Line_ORBCOMM_2018} and validated by scientific studies which primarily use the primary lobe \citep[e.g.][]{Hurley-Walker_leakage_2017} . The effort to recover the ``missing'' power led to the creation of a global gain calibration scheme.

It was observed that RF Explorers begin to leave their linear amplification zone at around -45 dBm\footnote{dBm - physical units of power, measured with respect to 1 milliwatt} and were definitely non-linear by -35dBm, where slices of the FEE model had visibly diverged from raw tile data (see Fig. \ref{fig:rfe-gain}). We begin by considering deformed AUT power P$_\mathrm{def}$, non deformed reference power P$_\mathrm{ref}$ and slices of the FEKO reference B$_\mathrm{ref}$ and FEE MWA beam B$_\mathrm{FEE}$ models for a satellite pass. Once Equation \ref{eq:beam_eq} is computed, information regarding absolute power recorded by the RF explorers is lost in favour of a normalized beam profile (see Section~\ref{ssec:feko}). Thus, gain calibration of the RF explorers must take place at the tile power level, before scaling or normalization processes distort the original power levels. A mask M$_\mathrm{def}$ is created using the region where the deformed tile power P$_\mathrm{def}$ exceeds -35dBm. This mask prevents the distorted sections of the measured primary beam from biasing the results of the multiple least-squares gain fits described below. 

Equation \ref{eq:beam_eq} is used to compute the deformed beam slice B$_\mathrm{def}$ using P$_\mathrm{def}$, P$_\mathrm{ref}$ and B$_\mathrm{ref}$. To maintain the initial power level, we mask the deformed section of  B$_\mathrm{def}$ using the mask M$_\mathrm{def}$ and use a least-squared method to determine a single multiplicative gain factor which will scale B$_\mathrm{def}$ down to the initial power level of P$_\mathrm{def}$. A similar method is used to scale the slice of the FEE beam B$_\mathrm{FEE}$ down to the initial power level of P$_\mathrm{def}$. The result of the scaling can be seen in Figure \ref{fig:rfe-gain} where B$_\mathrm{FEE}$ (yellow) and B$_\mathrm{def}$ (light blue) have been successfully scaled to match at low powers while clearly displaying a deficit of power at the peak of the primary beam. 

\begin{figure}
	\includegraphics[width=\columnwidth]{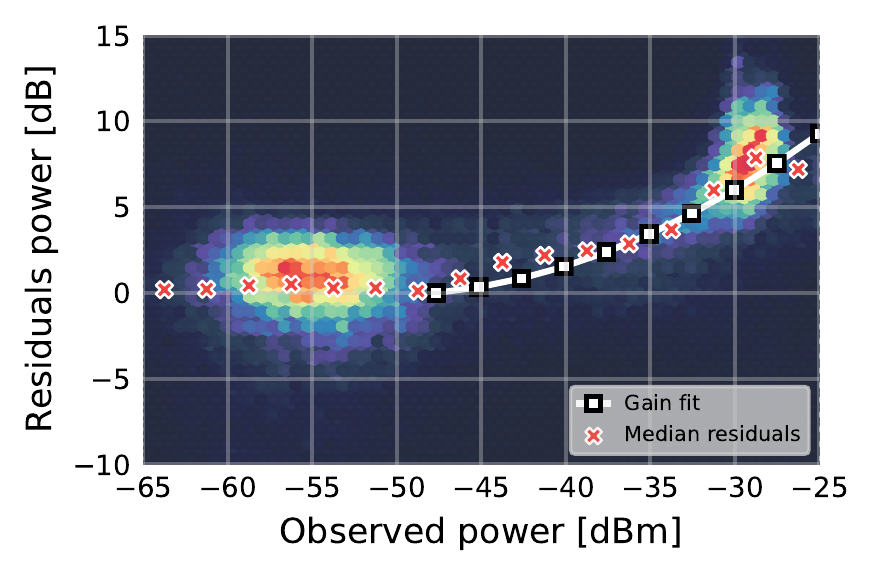}
    \caption{The 2D histogram distribution of high power distortions to RF signals. Ideally the residuals should have a value of 0 at all observed powers, indicating that the RF explorers reproduce input signals faithfully. The white curve is a 3$^{rd}$ order polynomial fit to the median values (black squares) of the data binned in $\sim$ 4dBm intervals.}
    \label{fig:rfe-gain-fit}
\end{figure}

We can now empirically determine a gain calibration solution by looking at the residual power (B$_\mathrm{FEE}$ - B$_\mathrm{def}$) of all satellite passes. The 2D histogram of all residual power is shown in Figure \ref{fig:rfe-gain-fit}, with the horizontal axis representing power observed by the AUT RF explorers, and the vertical axis representing residual power. The figure displays a bridged bimodal distribution, which can be explained by considering the profile shape of cross sectional slices of the MWA beam models. The nodes at the edges of the primary beam are sharply peaked and extremely narrow, leading to a dearth of observational data points in such regions as satellites pass over them relatively quickly. The cluster of points at lower observed power is the result of satellites passing over the relatively broad secondary lobes of the MWA beam while the cluster at higher observed power comes from satellite passes transiting through the primary beam. For linear gain internal to the RF Explorer, one would expect the residuals to be $\sim$ zero, while positive residuals result from non linear gains. The white curve and associated black squares are a 3$^\textrm{rd}$ order polynomial fit to the median values (red crosses) of the data binned in $\sim$ 4dBm intervals. This clearly demonstrates that the RF explorers gradually enter the non-linear regime at $\sim -40$dBm and exhibit residuals of $\sim6$dB at observed powers of --30 dBm. 

The result of applying the calibration solution developed above to a single satellite pass are seen in Figure \ref{fig:rfe-gain} where B$_\mathrm{def}$ (light blue) is scaled up to B$_\mathrm{cali}$ (navy blue) and represents a much better fit to a slice of the fee model B$_\mathrm{FEE}$ (yellow).

The RF Explorer gain calibration technique presented in this section has been shown to be necessary but comes with a minor drawbacks. Primarily, the global nature of our method could result in the loss of potentially interesting structure present at the center of the primary lobe. The accuracy of the primary lobe has been validated by multiple studies \citep[e.g.][]{Line_ORBCOMM_2018, Hurley-Walker_leakage_2017} and deviations are not expected. The gain correction was essential as the absolute scale of fluctuations in the more uncertain side-lobes, were determined by fitting satellite signals to the well characterized primary lobe. These corrections also enable us to regenerate all-sky beam maps which may be utilized in further studies. Future iterations of this experiment, which could be scaled to passively monitor the full MWA array or SKA-Low, will have to extensively characterize off-the-shelf components such as RF Explorers. Our characterization of the gain profile revealed that the accuracy of factory specification may not be sufficient for experiments of such sensitivity and scale.

\subsection{Tile maps}
\label{ssec:tile-map}

We now create MWA beam maps using the method described in Section \ref{sec:data} with the caveat that the RF explorers gain calibration solution described in Section \ref{ssec:rfe-gain} are applied to all data from the AUT RF explorers. A single multiplicative gain factor, determined by least-squares minimization, is used to scale measurements to the level of the zenith-normalized FEE beam model. Before B$_\mathrm{AUT}$ can be projected onto a HEALPix map, it must pass a final goodness-of-fit test. The frequency mapping method described in Section \ref{ssec:freq_map} has been highly successful at dealing with the massive volume of data produced over the course of this experiment, but does exhibit an $\sim2\%$ failure rate, where the transmission frequency of satellites is misidentified. To catch these final outliers, a chi-squared p-value goodness-of-fit test between the scaled measured beam B$_\mathrm{AUT}$ and the FEE model B$_\mathrm{FEE}$ is implemented, with a threshold tuned to ensure that only the beam profiles with obviously wrong null positions are rejected. Successful satellite passes are projected onto a HEALPix map representative of an accurate all-sky MWA beam response. 

\begin{figure*}
	\includegraphics[width=\textwidth]{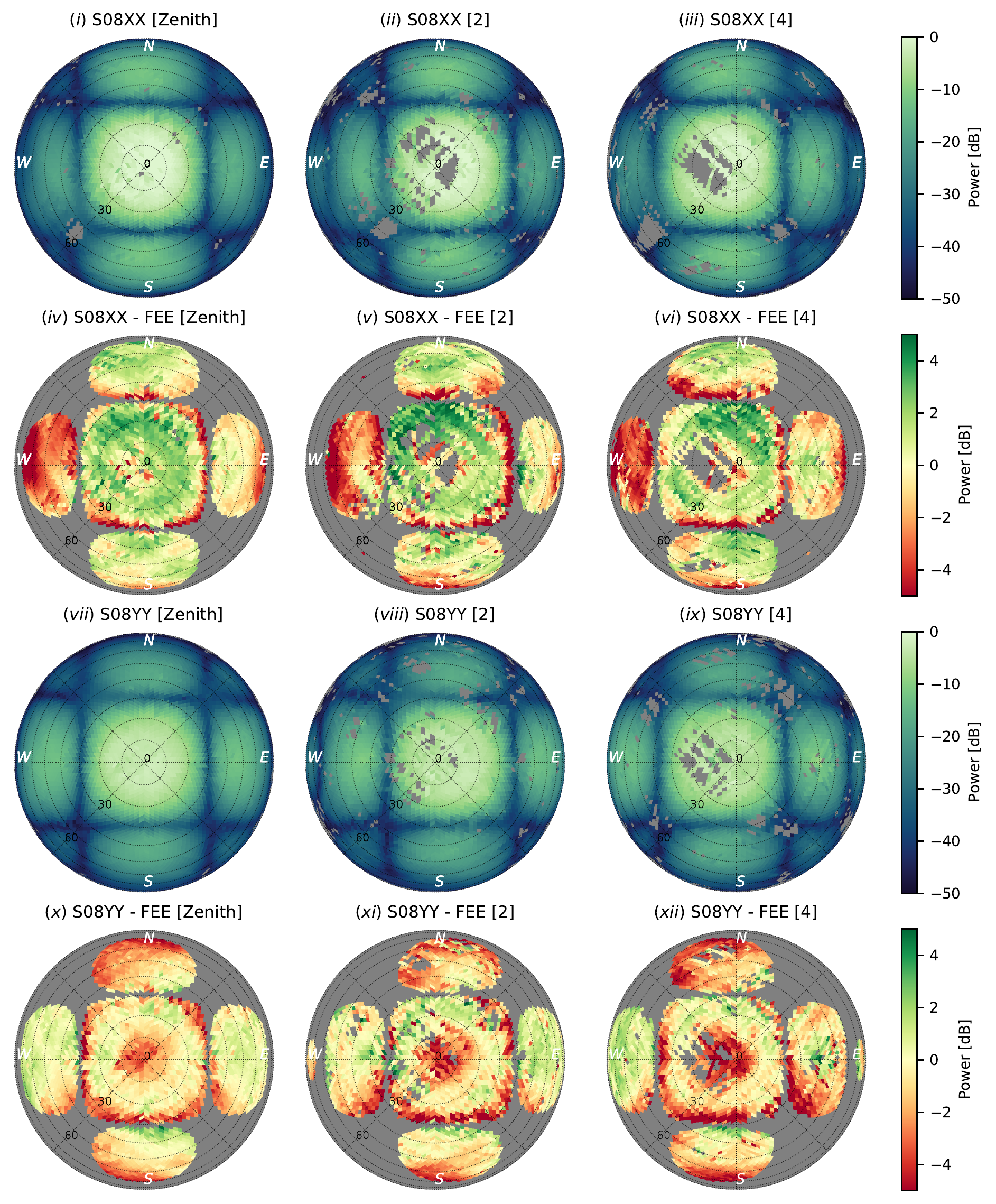}
    \caption{A set of beam maps measured for tile S08. The first row (i)-(iii) are maps of the XX polarization of tile S08, while the second row (iv)-(vi) represent the ratios between the beam maps and the corresponding FEE models. The three columns represent maps at the zenith, 2 and 4 pointing of the MWA. The last two rows (vii)-(xii) are an identical analysis for the YY polarization of tile S08.}
    \label{fig:tile-grid}
\end{figure*}

A set of tile maps at multiple pointings and polarisations are shown in Figure \ref{fig:tile-grid}, created with data from tile S08 and Ref$_1$. The residual maps shown in the second and fourth row of Figure \ref{fig:tile-grid} display large gradients in power at the Southern and Eastern edges of their primary lobes. This effect is attributed to gradients in the ground screen of the MWA tiles. Such gradient can lead to systematic angular offsets from the intended pointing of the MWA tiles specified by the beamformers. This effect is most pronounced at the steep nulls surrounding the primary lobe where systematic displacements in null positions occur. The mismatch in the measured position of the nulls as compared to the FEE model manifest as the gradients observed in the residual maps of Figure \ref{fig:tile-grid}.

\begin{figure}
	\includegraphics[width=\columnwidth]{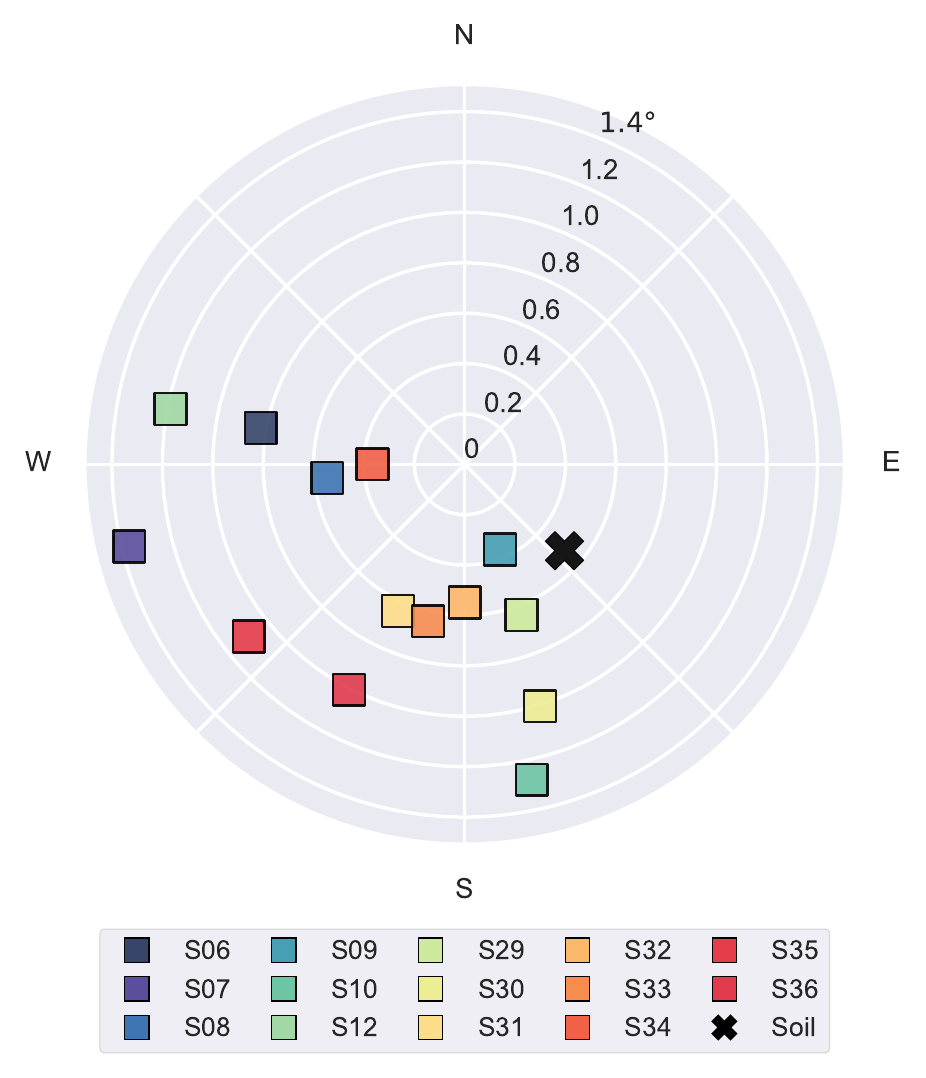}
    \caption{Measured angular offsets of zenith beam maps. The black cross represents a pervasive $\sim0.5^\circ$ gradient of the soil in the Southern Hex towards the South-East.}
    \label{fig:tilt}
\end{figure}

We further investigate this effect to determine the gradient of the ground screens of our MWA tiles. This is achieved by displacing our measured beam maps and minimizing the residual power at the edge of the primary lobe. The gradient in the ground screens were determined to $\sim15$ arcmin resolution by interpolating our HEALPix maps to a higher resolution with NSIDE=256. Figure \ref{fig:tilt} shows the measured angular offset of our 14 tiles from the zenith pointing. Local surveys of the tiles in the Southern Hex have identified a gradual half degree gradient in the soil, from the North-West to the South-East which would result in all beams being offset by $\sim0.5^\circ$ towards the South-East, displayed as the black cross in Figure \ref{fig:tilt}. This analysis shows a significant scatter in angular beam offsets, indicative of tile gradients ranging up to 1.4$^\circ$, and vertical displacements exceeding 10 cm over a 5 m ground screen.  

\begin{figure*}
	\includegraphics[width=\textwidth]{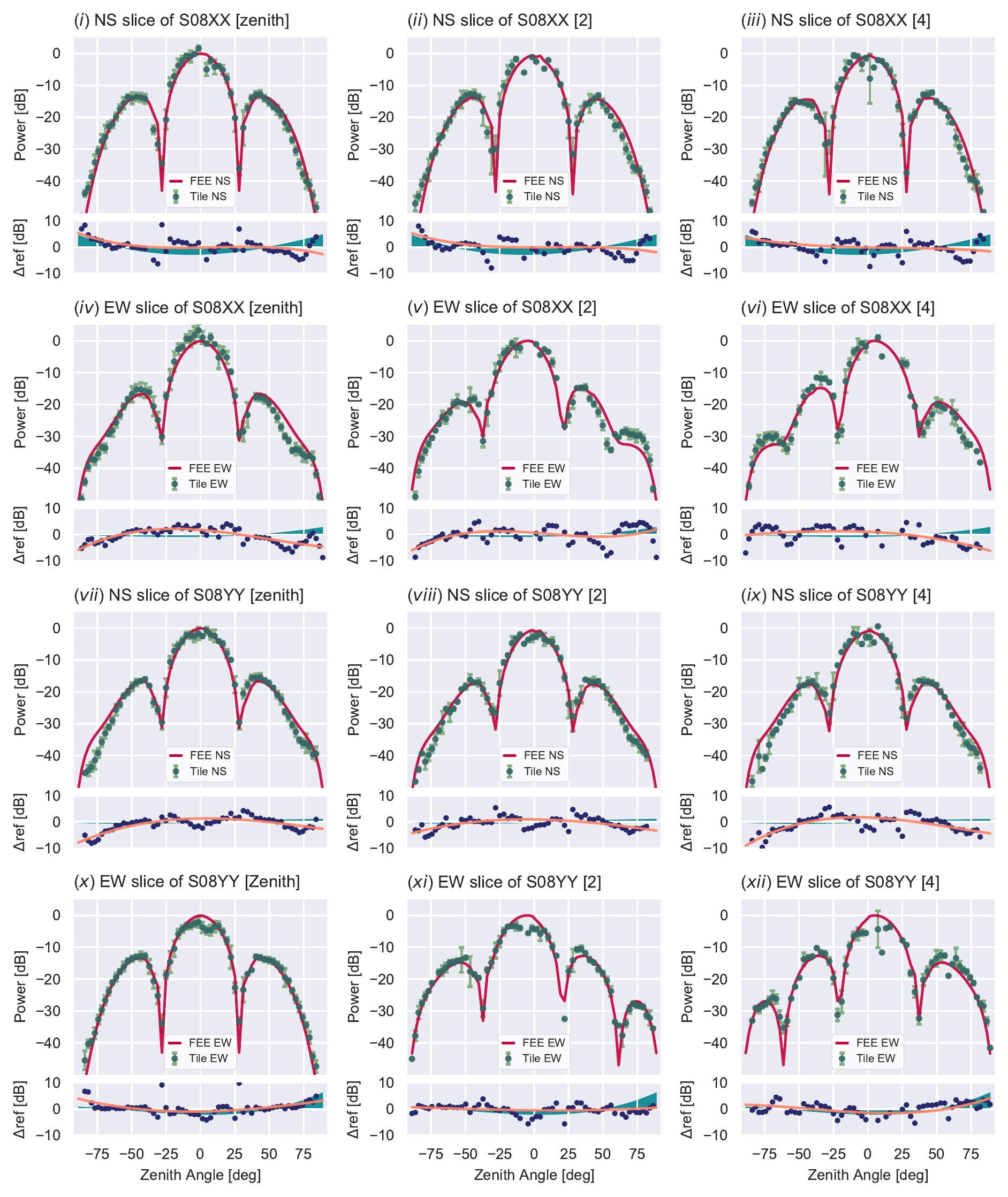}
    \caption{North-South(NS) and East-West(EW) slices of beam maps (S08) presented in Figure. \ref{fig:tile-grid}. The first row (i)-(iii) displays NS slices of tile S08 compared to corresponding slices of the FEE model, in the XX polarization and at three pointings. The lower panels show the residuals between the measured tile maps and the FEE models, with the cyan shaded regions representing errors which can be attributed to the reference antennas. The second row (iv)-(vi) display EW slices of S08XX. The bottom two rows (vii)-(xii) represent an identical analysis for the YY polarization of tile S08.}
    \label{fig:tile-slices}
\end{figure*}

In Figure \ref{fig:tile-slices}, NS and EW slices of the tile maps are compared to the corresponding FEE models. The first row (subplots (i)-(iii)) represent NS slices of the XX beam map of tile S08. The lower panels of these subplots explore residual power between measurements and the FEE model. The orange curve represents a 3$^{\textrm{rd}}$ order polynomial fit to the residual power, while the cyan shaded regions account for errors which can be attributed to the reference tiles, as seen from the null tests (See \secref{ssec:null-test} and Fig. \ref{fig:null-test}). The second row (subplots (iv)-(vi)) represents an identical analysis for the EW slice of the beam maps. The last two rows (subplots (vii)-(xii)) complete the analysis described above, for the YY beam maps. The three columns represent the zenith and the 2 and 4 off-zenith pointings.

\begin{figure*}
	\includegraphics[width=\textwidth]{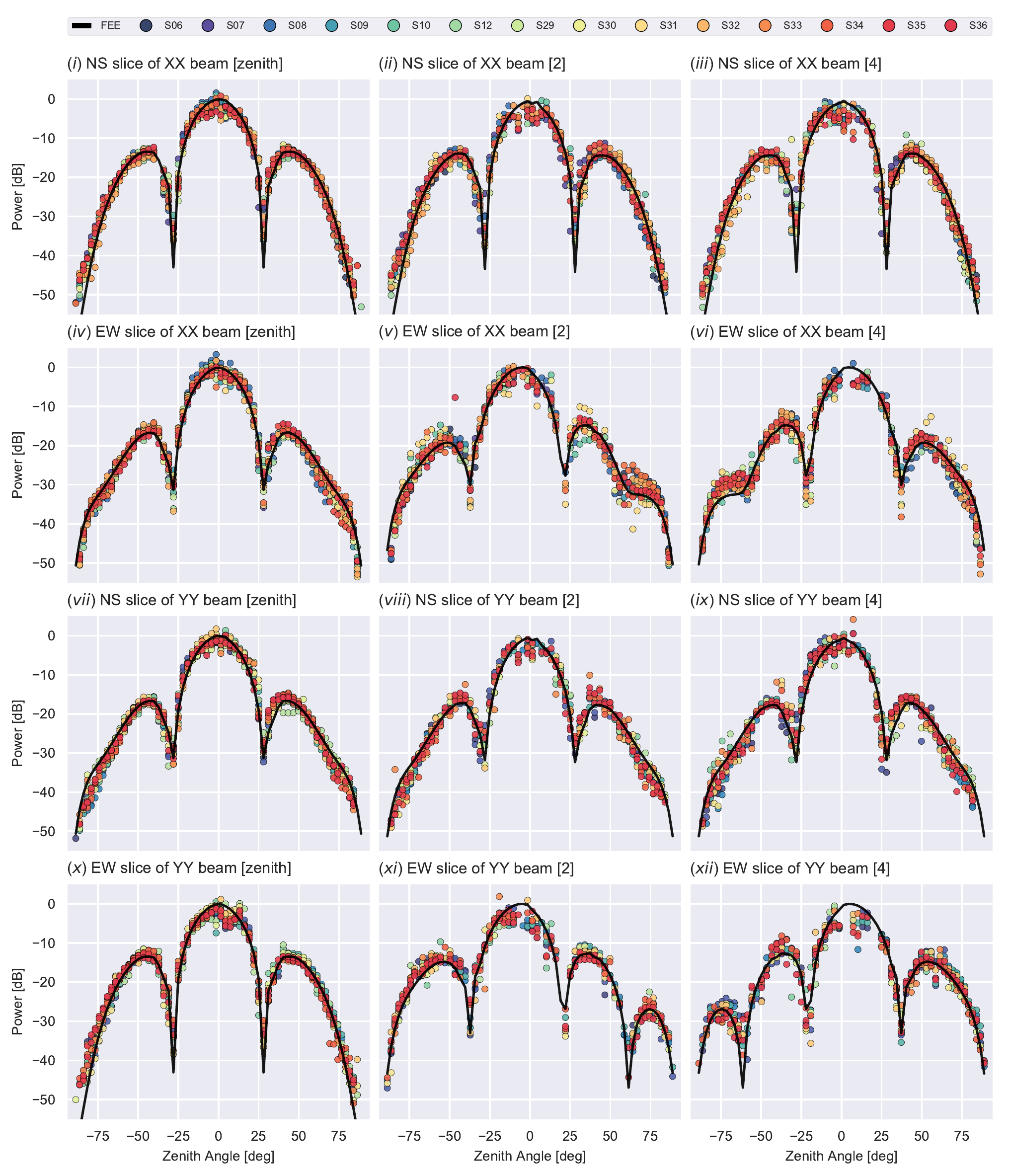}
    \caption{Distribution of all beam maps compared to corresponding slices of the FEE beam. The three vertical columns represent the Zenith, 2, 4 MWA pointings, while the horizontal rows represent cardinal (NS, EW) slices of beam maps at both polarizations (XX, YY).}
    \label{fig:beam-var}
\end{figure*}

The distribution of beam shapes of our 14 tiles can been seen in Figure \ref{fig:beam-var}, displayed as cross-sections of the beam maps along the cardinal axes. The marginally larger scatter observed of data points around the primary lobes can be attributed the global RF Explorer gain calibration described in \ref{ssec:rfe-gain}.

A subtle but interesting pattern emerges from Figures \ref{fig:tile-slices} and \ref{fig:beam-var}. Consider the first row of subplots, representing NS slices of XX beams. There is a slight excess of measured power ($\sim$2dB) at the outer edge of the secondary lobes. XX dipoles are EW oriented and are most sensitive perpendicular to their physical orientation. Thus our measurements indicate a greater than expected sensitivity along the most sensitive axis of the XX dipole. Similarly, the YY beam oriented along the NS, measures an excess of power along its most sensitive axis (EW) as seen in the fourth rows of Figures \ref{fig:tile-slices} and \ref{fig:beam-var}. Conversely, the power measured by the dipoles along their least sensitive axis, parallel to their physical orientation, is less than expected. This is seen in the second and third rows of Figures \ref{fig:tile-slices} and \ref{fig:beam-var}.

\begin{figure}
	\includegraphics[width=\columnwidth]{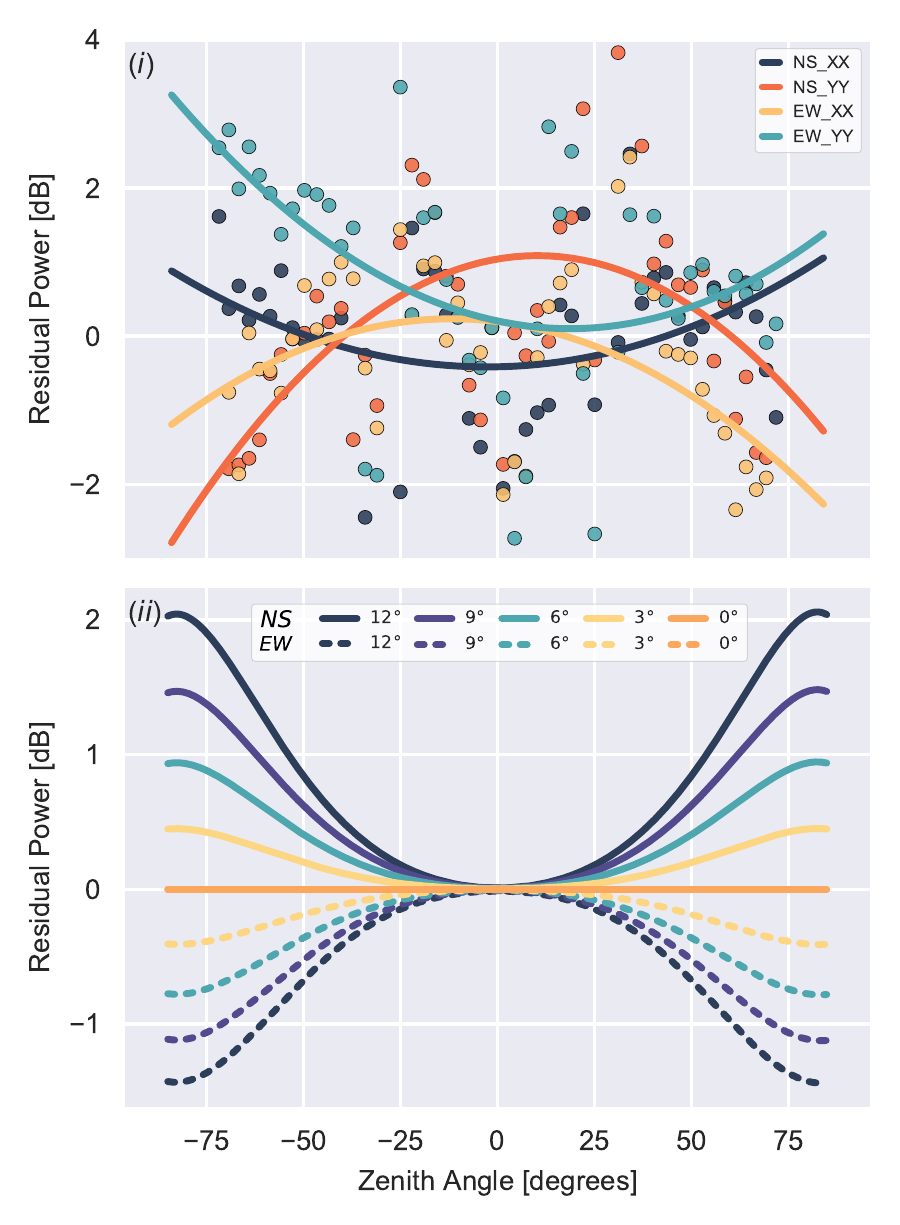}
    \caption{(\textit{i}) Global residual power averaged over all 14 tiles. Each point represents the median power at various zenith angles, along cross-sectional slices of residual power between measured MWA tile maps and the FEE model. Second order polynomials fit to this data reveal systematic offsets attributed to rotations in the reference tiles. (\textit{ii}) Simulated effect of anticlockwise rotation in the XX reference tile on measurements of MWA tile power. The solid and dashed lines represent North-South and East-West cross-sectional slices of the beam model.}
    \label{fig:rot_resi}
\end{figure}

This effect was investigated by computing median residual power for all 14 tiles along EW and NS slices. The results shown in Figure \ref{fig:rot_resi} (\textit{i}) have been fit with a 2$^{\textrm{nd}}$ order polynomial which shows systematic, radially dependent offsets. This residual structure could potentially be attributed to a rotation of the reference antenna. This scenario was explored using the simulated FEKO models of the reference, as shown in Figure \ref{fig:rot_resi} (\textit{ii}). The solid lines represent the residual power which would be measured along a NS slice of the XX MWA beam if the reference tile was rotated by a range of angles, while the dashed lines represent the EW slices. As observed in Figures \ref{fig:tile-slices} and \ref{fig:beam-var}, the excess measured power along the most sensitive axis of the dipole and the deficit of power along the least sensitive axis of the dipoles seem to have similar shapes to the simulated residuals of reference antenna rotations. Unfortunately, the degeneracy inherent to the symmetric reference FEKO models prevents the identification of the direction of this rotation.

\begin{figure}
	\includegraphics[width=\columnwidth]{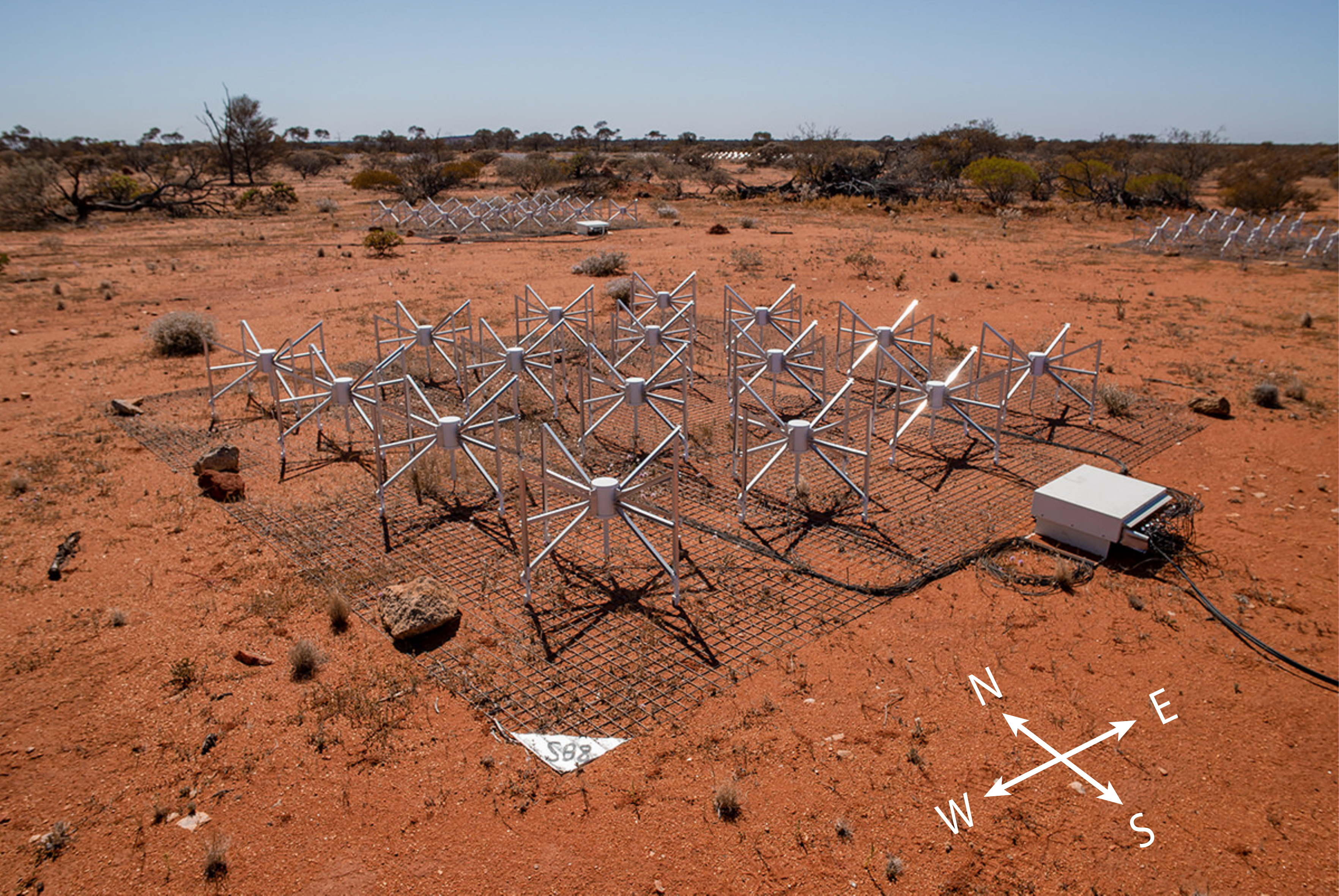}
    \caption{A current image of the condition of tile S08 reveals several large rocks on the western edge of the ground screen as well as a significant amount of loose soil which has covered portions of the metal mesh of the ground screen.}
    \label{fig:S08}
\end{figure}

The measurements of the western sidelobe of tile S08 show a significant deficit in power of order $\sim$4 dB, seen in the second row of Figure \ref{fig:tile-grid}. Pictures of the in-situ condition of the tile reveal potential environmental factors which could potentially be responsible. In Figure \ref{fig:S08}, we observe a number of large rocks on the ground screen of the tile. Additionally, the harsh weather conditions on site seem to have swept some loose soil onto the ground screen, partially obscuring the metal mesh. Both these effects are most prominent along the western edge of the tile and could plausibly explain the measured deficit of power. 

\section{Conclusions}
\label{sec:conclusions}

We have measured the all-sky beam response of 14 onsite MWA tiles, at both instrumental polarizations and at three pointings. As the first dual polarization MWA beam measurement experiment, both our XX and YY beam maps display good agreement with the cutting-edge FEE beam models \citep{Sokolowski_FEE_2017} to first order. Further investigations reveal a range of environmental perturbations from the FEE models, which may present the scope for improved calibration for EoR and other science cases. 

The most significant distortions to the MWA beams are found to be asymmetric and in one or more of the sidelobes. These distortions have been observed to occur due to environmental effects such as the obscuration of the metal mesh of the ground screen by loose soil, and other large objects such rocks (see Figure \ref{fig:S08}). Further, local foliage surrounding the tile may also contribute to beam deformations in an unpredictable, non-static manner as they grow and wither over the course of a year. These effect have been seen to deform the sidelobes with up to a $\sim5$dB deficit in measured power. Such effects are at the level of $\sim10\%$ zenith power and could have a serious effect on multiple science cases. 

Further, investigations into the mismatch of positions of the primary nulls have revealed the effects of gradients in soil and ground screens. The Southern Hex is known to have a 0.5$^\circ$ gradient in the soil from the NW to the SE. Our investigation revealed the existence of local soil gradients, beyond the gradual background gradient, up to $\sim1.5^\circ$, scattered around the local gradient (see Figure \ref{fig:tilt}). This effect results in the bore-sight of the MWA beam pointing slightly away from its expected position, and is analogous to pointing errors in traditional telescopes. While these effects do not significantly effect the central portion of the primary lobe, the steep edges of the beam surrounding the nulls are susceptible to large power offsets, approaching $\sim8$dB, with pointing offsets as low as $\sim1.5^\circ$. This may be of particular import to observations conducted during the day, where clever observation techniques are used to place the sun in one of the primary beam nulls and achieve maximum attenuation \citep{Morgan_IPS_2019}. Positional offsets of the nulls could potentially introduce significant erroneous solar flux to such observations.

Finally, unexpected deviations from measured power, along and perpendicular to dipole axis have revealed rotations in our reference tiles. Degeneracies resulting from symmetries in the reference antenna beam models have prevented the exact identification of the direction of this rotation. While the rotation of the reference tile does not affect MWA science cases, it does present an interesting proxy to study effects rotations in MWA tiles may have. MWA tiles are aligned to within 0.5-1.0$^\circ$ of the NS meridian. We estimate that such uncertainties in rotation may introduce error less than $\sim1$dB close to the horizon. Such effects do not significantly effect the primary lobe but increase radially outwards. Interferometric arrays such as the MWA are prone to flux leakage, which may be exacerbated by rotational errors which will increase the coupling between the orthogonal, independent dipoles. With better calibration of reference antennas, future versions of this satellite experiments will be able to measure rotations of MWA beams, enabling studies of the effects of tile rotations and place upper limits on the acceptable leeway in beam rotations. More comprehensive simulations of such effects have been reserved for future works.

These measurements provide useful insight to various beam deformations, but are limited to an extremely narrow frequency band. This is an apparent shortcoming of satellite based beam measurement techniques and presents a significant impediment to the adoption of satellite based beam maps by the radio astronomy community. So far, the utility of such methods has been limited to the validation of advanced electromagnetic simulations. Future work based on our measured beam maps will investigate the fitting of 32 complex-valued gain parameters of the FEE model \citep{Sokolowski_FEE_2017} to create perturbed FEE models representative of measurements at 137 MHz. Future investigations will explore the efficacy of extrapolating these gain values to cover the MWA's frequency band, potentially opening an avenue for broadband, pseudo-realistic beam models. Finally, a study combining our measured beam maps and data from the regular short dipole tests, used to find dead dipoles, may enable the creation of more realistic beam models for use with archival MWA data.  

The implications of more accurate beam models are far-reaching. The detection of the EoR and studies of the cosmic dawn are key science cases of the MWA and upcoming telescopes such as the SKA-Low. The extreme dynamic range of such experiments necessitate uncompromising precision, which may be impeded by imperfect beam models. Particularly, imperfect beam models with radially increasing uncertainty can result in flux calibration errors of bright sources such as Fornax A, and the diffuse galactic plane, close to the horizon. The intrinsic chromatic nature of radio interferometers can be exacerbated by in-situ beam distortions. Such effects could lead to the introduction of bright, unphysical spectral structure, impeding the detection of the EoR signal \citep[e.g.][]{Byrne_redundant_2019, Orosz_2019}.

Large precise surveys such as GLEAM \citep{Hurley-Walker_leakage_2017}, along with planned surveys such as GLEAM-X and LoBES, use the half-power portion of the central lobe. This has primarily been necessary due to beam modelling errors, and results in a loss of sensitivity and survey efficiency. Instruments such as the MWA are sensitive to large portions of the sky, and approach all-sky sensitivity at  low frequencies. Increased confidence in beam models would enable larger swatches of the sky to be observed at a time. Additionally, unresolved sources in the sidelobes contribute to confusion-noise, place lower limits on sensitivity to faint sources. 

This experiment has demonstrated the feasibility of a passive parallel monitoring system, built from off-the-shelf and relatively inexpensive components, which could easily be scaled up to monitor the beam shapes of the entire MWA array, providing invaluable information to many science cases and improving calibration across the board. Using individual tile models for calibration of MWA data is possible using pipelines such as the RTS and FHD, but presents a significant increase in computational effort. The level of measured beam distortions and their complex nature reinforces our conjecture that more realistic beam shapes could significantly improve the accuracy and sensitivity of science possible using the MWA. Further investigation and detailed simulations will be necessary to understand how realistic beam models will effect calibration and improve results. If successful in improving calibration, a similar passive parallel monitoring system may be an essential tool for upcoming telescopes. This is particularly applicable to SKA-Low stations constructed of 512 dipoles, with multiple degrees of freedom available for complex beam perturbations.

This experiment has led to the development of an open-source \texttt{python} package called EMBERS\footnote{\url{https://embers.readthedocs.io}} -- Experimental Measurement of BEam Responses with Satellites. EMBERS is almost completely parallelized and is capable of being scaled to much larger arrays, enabling an end-to-end analysis of satellite beam measurement data. EMBERS can be used and modified by anyone, with the aim of enabling the measurement of beam shapes of radio telescopes all over the world with ease. 

\section{Data Availability}

The final data products of this project - dual polarization beam maps of 14 MWA tiles at multiple pointings along with a set of diagnostic plots - are available at \url{https://github.com/amanchokshi/MWA-Satellite-Beam-Maps}. On request, the complete raw data set can be made available. This includes over 4000 hours of RF satellite data which can be used to recreate all results in this paper using the EMBERS packages. 

\section*{Acknowledgements}

We would like to thank the MWA operations team for their constant support and invaluable engineering expertise. This research was supported by the Australian Research Council Centre of Excellence for All Sky Astrophysics in 3 Dimensions (ASTRO 3D), through project number CE170100013. This scientific work makes use of the Murchison Radio-Astronomy Observatory, operated by CSIRO. We acknowledge the Wajarri Yamatji people as the traditional owners of the Observatory site. 




\bibliographystyle{mnras}
\bibliography{citations} 







\bsp	
\label{lastpage}
\end{document}